\documentclass[10pt,a4paper]{elsarticle} 
\usepackage{amsmath,amssymb,setspace,tikz,longtable,url,pgfplots}
\pgfplotsset{compat=newest}
\usepackage{natbib}
\usepackage{thmtools}
\usepackage{tasks}
\usetikzlibrary{arrows}
\newtheorem{case}{Case}
\begin{document}

\title{On the Kolkata index as a measure of income inequality}
\author{Suchismita Banerjee}
\address{Economic Research
	Unit, Indian Statistical Institute, Kolkata,
	India. \\ Email: suchib.1993@gmail.com}

      \author{Bikas K. Chakrabarti} \address{Saha
        Institute of Nuclear Physics, Kolkata, and \\ 
        Economic Research Unit, Indian Statistical Institute, Kolkata, India. \\
    Email: bikask.chakrabarti@saha.ac.in}

      \author{Manipushpak Mitra} \address{Economics Research
        Unit, Indian Statistical Institute, Kolkata,
        India. \\
    Email: mmitra@isical.ac.in}

      \author{Suresh Mutuswami} \address{School of Business, University of Leicester,
        Leicester, United
        Kingdom, and \\        
        Economics Research
        Unit, Indian Statistical Institute, Kolkata,
        India.\\ Email: smutuswami@gmail.com}
	

\begin{abstract}
	We study the mathematical and economic structure of the Kolkata ($k$) index of income inequality. We show that the $k$-index always exists and is a unique fixed point of the complementary Lorenz function, where the Lorenz function itself gives the fraction of cumulative income possessed by the cumulative fraction of population (when arranged from poorer to richer). We argue in what sense the $k$-index generalizes Pareto's 80/20 rule. Although the $k$ and Pietra indices both split the society into two groups,  we show that $k$-index is a more intensive measure for the poor-rich split. We compare the normalized $k$-index with the Gini coefficient and the Pietra index and discuss when they coincide.  Specifically, we identify the complete family of Lorenz functions for which the three indices coincide. While the Gini coefficient and the Pietra index are affected by transfers exclusively among the rich or among the poor, the $k$-index is only affected by transfers across the two groups. 
\end{abstract}  


\maketitle

\section{\textbf{Introduction}}
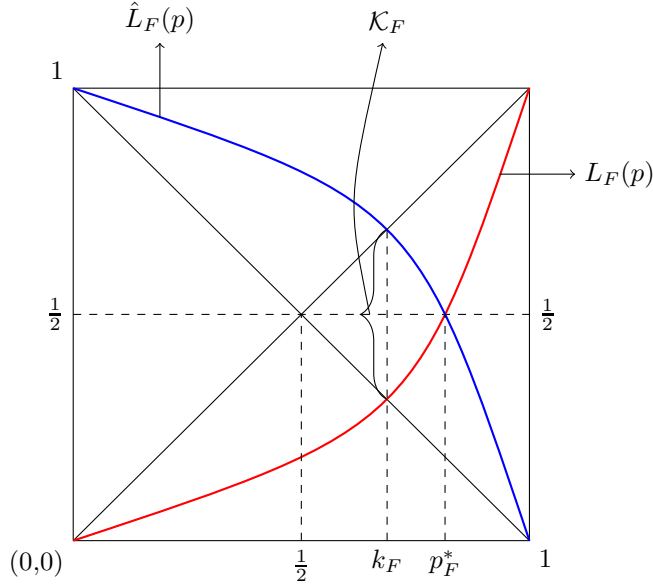
\begin{figure}[h]
	\begin{center}
	\begin{tikzpicture}[scale=6]
	\draw (0,0)--(0,1)--(1,1)--(1,0)--(0,0);
	\draw (0,0)--(1,1) node [pos=0,below left] {(0,0)};  
        \draw (0,1)--(1,0)node [pos=0,above left] {1} node
        [pos=1,below right] {1}; \draw [dashed] (0,0.5)--(1,0.5)node
        [pos=0,left] {$\frac{1}{2}$} node [pos=1,right]
        {$\frac{1}{2}$}; \draw [red,thick]
        (0,0)..controls(0.75,0.25)..(1,1); \draw [blue,thick]
        (0,1)..controls(0.75,0.75)..(1,0); \draw [dashed]
        (0.688,0)--(0.688,0.688)node [pos=0,below] {$k_F$}; \draw
        [dashed] (0.815,0)--(0.815,0.5)node [pos=0,below] (p*) at
        (0.815,-0.12) {$p^*_F$};
	\draw [dashed] (0.5,0)--(0.5,0.5) node [pos=0,below] {$\frac{1}{2}$};
	\draw [->] (0.935,0.81)--(1.1,0.81)node [pos=1,right] {$L_F(p)$};
	\draw [->] (0.19,0.935)--(0.19,1.1)node [pos=1,above] {$\hat{L}_F(p)$};
	\draw [decorate,decoration={brace,amplitude=10pt},xshift=0pt,yshift=0pt]
	(0.688,0.312)--(0.688,0.688);	\node (k) at (0.6888,1.1) [above]
	{$\mathcal{K}_F$};
	\draw [->,black] (0.65,0.5)..controls(0.6,0.75)..(k);
	\end{tikzpicture}
	\caption{The red curve is the Lorenz function and the blue
          curve is the complementary Lorenz function. The normalized
          $k$-index is $\mathcal{K}_F=k_F-L_F(k_F)=2k_F-1$ and
          $p^*_F=L^{-1}_F(\frac{1}{2})$ is the population proportion
          associated with the point of intersection of the Lorenz and
          the reverse order Lorenz functions (colour
          online). \label{fig:lorenz}}
      \end{center}
      \end{figure}

\noindent In the Lorenz curve (see \cite{Lo1905} for more details) one plots the proportion of the total income of the
population that is earned by the bottom $p$ proportion of the population. See Figure \ref{fig:lorenz}, where we plot accumulated proportions of the population from poorest to richest along the horizontal axis and total income held by these proportions of the population along the vertical axis.  The $45^\circ$ line represents a situation of perfect equality.

Often we are interested in a summary statistic of the Lorenz function.\footnote{We will use the terms Lorenz function and Lorenz curve interchangeably in this paper.} This is because the Lorenz curves can intersect each other meaning that we
cannot order the curves.  One way of dealing with this is to rely on a summary statistic (see \cite{Aa2000} for more details).  The most popular summary statistic is the \emph{Gini coefficient} (see \cite{Gi1912}) which is the ratio of the area between the $45^\circ$ line and the Lorenz curve to the total area under the $45^\circ$ line. The \emph{Pietra index} (see \cite{Pi1915}) is the maximum value of the gap between the $45^\circ$ line and the Lorenz curve (see also
\cite{ES2010}). 

In this paper, we are specifically interested in a particular summary index called the \emph{Kolkata index} or the $k$-index (see \cite{GC2014} for more details) which is that proportion $k_F$ such that $k_F + L_F(k_F) = 1$ where $L_F(p)$ is the Lorenz function  (also see \cite{CG2017}).  We can understand $k_F$ better as follows.  Suppose we split society into two groups: the ``poor'' who constitute a fraction $p$ of the population and the remaining ``rich.''  Note that $L_F(p) \leq p$; hence, $p$ is an upper bound on the income share of the poor.  The actual share of the rich,
on the other hand, is $1 - L_F(p)$.  The $k$-index splits society into two groups in a way that the \emph{egalitarian income share} of the
poor equals the \emph{actual income share} of the rich.\footnote{\cite{Su2010} uses the $k$-index to define a generalized Gini coefficient.}

      The $k$-index takes values in the range $[1/2, 1]$ which makes
      it different from the Gini coefficient and the Pietra index,
      both of which take values in the range $[0, 1]$.  However, a
      simple normalization of the $k$-index, namely
      $\mathcal{K}_F \equiv 2k_F-1$, achieves this.  Like the other
      two indices, the extreme values of the normalized $k$-index
      correspond to complete equality ($\mathcal{K}_F = 0$) and
      complete inequality ($\mathcal{K}_F = 1$) respectively. The normalized $k$-index was first introduced in \cite{EL2015A} and was called the ``perpendicular-diameter index'' (see \cite{EL2015A}, \cite{EL2015B}, \cite{EL2016}).

      We show that the $k$-index is a fixed-point of the function
      $\hat{L}_F(p)\equiv 1 - L_F(p)$ which we call the
      \emph{complimentary Lorenz function}.  In particular, we show
      that the fixed-point exists and is unique for all Lorenz
      functions.  We also show that the $k$-index generalizes Pareto's
      $80/20$ rule: ``$20$\% of the people own $80$\% of the income.''
      The $k$-index has the property that $[100(1 - k_F)]$\% of the
      people own $[100k_F]$\% of the income.  Or, equivalently,
      $[100k_F]$\% of the people only have $[100(1 - k_F)]$\% of the
      income.  We show that both the $k$ and the Pietra indices split
      society into two groups and we discuss the differences between
      the two indices in this regard. We compare the
      normalized $k$-index with the Gini coeffeicient and the Pietra
      index and obtain certain important conclusion in terms of
      coincidence possibilities between all or any two of these three
      measures.  We show that for any given income distribution the value of Gini coefficient is no less than that of the Pietra index and the value of the Pietra index is no less than that of the normalized $k$-index. We also  demonstrate that while the Gini coefficient and the Pietra index are affected by transfers exclusively among the rich or among the poor, the $k$-index ranks is only affected by transfers across the two groups.  

\section{\textbf{The framework}}\label{sec2}
Let $F$ be the distribution function of a non-negative random variable
$X$ which represents the income distribution in a society.  The
\emph{left-inverse} of $F$ is defined as
$F^{-1}(t) = \inf_{x}\{x |F(x) \leq t\}$.  
Assume that the mean income $\mu = \int_{0}^{\infty}xdF(x)$
is finite.  In this case, we obtain an alternative representation of
the mean: $\mu = \int_{0}^{1}F^{-1}(t)dt$.  

The \emph{Lorenz function}, defined as
$L_F(p) =(1/\mu)\int_{0}^{p}F^{-1}(t)dt$, gives the proportion of
total income earned by the bottom $100p\%$ of the population.  The
following properties of the Lorenz function are well-known: (i)
$L_F(0) = 0, L_F(1) = 1$ and $L_F(p) \leq p$ for $p \in (0, 1)$, and
(ii) the Lorenz function is continuous (see \cite{Ga1971}),
non-decreasing and convex.

The \emph{complementary Lorenz function} is defined as
$\hat{L}_F(p) = 1 - L_F(p)$ . It measures the proportion of the total
income that is earned by the top $100(1 - p)\%$ of the population:

\begin{equation}
  \hat{L}_F(p):=1-L_F(p)=1-\frac{\int\limits_{0}^{p}F^{-1}(t)dt}{\mu}=\frac{\int\limits_{p}^{1}F^{-1}(t)dt}{\mu}.
\end{equation}

It follows straightforwardly that
$\hat{L}_F(0) = 1, \hat{L}_F(1) = 0$, and $0 \leq \hat{L}_F(p) \leq 1$
for $p \in (0, 1)$.  Furthermore, $\hat{L}_F$ is continuous, non-increasing and
concave on $(0, 1)$.

The \emph{Gini coefficient} is given by
$G_F = 2\int_{0}^{1}(p - L_F(p))dp$. The \emph{Pietra index} maximizes
$p - L_F(p)$.  It is easy to show that this function is maximized at
$p = F(\mu)$; hence, we can define the Pietra index alternatively as
$\mathcal{P}_F = F(\mu) -L_F(F(\mu))$. Different representations of these indices can be found in \cite{EL2018}. 


\section{\textbf{Structure of the $k$-index}}

\subsection{$k$-index as a fixed point of the complementary Lorenz function}
As mentioned, the $k$-index is defined by the solution to the equation
$k_F + L_F(k_F) = 1$.  It has been proposed as a measure of income
inequality (see \cite{CG2017}, \cite{GC2014} for more details). We can
rewrite $k_F + L_F(k_F) = 1$ as $k_F = 1 - L_F(k_F) =
\hat{L}_F(k_F)$. Hence, the $k$-index is a fixed point of the
complementary Lorenz function. Since the complementary Lorenz function
maps $[0, 1]$ to $[0, 1]$ and is continuous, it has a fixed point by
Brouwer's fixed point theorem.  Furthermore, since $\hat{L}_F(p)$ is
non-increasing, the fixed point has to be unique.

We can say a little more about the location of the fixed point.  Let
$p_F^* = L_F^{-1}(1/2)$. Given any Lorenz function $L_F(p)$, $p^*_F$ is that fraction of the population having 50\% of the total income and this fraction is intimately related to the ``horizontal-diameter index'' (see \cite{EL2015A}, \cite{EL2015B}, \cite{EL2016} and \cite{EL2018}).  Observe that $p_F^* \geq 1/2$ with the
equality holding only if we have an egalitarian income distribution. We
claim that the unique fixed point of $\hat{L}_F$ lies in the interval
$[1/2, p_F^*]$.  Let $Z_F(p) = \hat{L}_F(p) - p, p \in [0, 1]$. Note
that $Z_F$ is continuous.  Since $L_F(p) \leq p$, we have
$Z_F(1/2) \geq 0$.  Also,
$Z_F(p_F^*) = \hat{L}_F(p_F^*) - p_F^* \leq \hat{L}_F(p_F^*)
-L_F(p_F^*) = 0$. It follows from the Intermediate Value Theorem that
there exists $k_F\in [1/2, p_F^*]$ such that $Z_F(k_F) = 0$.
Therefore, we have established the following:
\begin{enumerate}
	\item[{\bf (FP)}] There exists a $k_F\in [1/2,p^*_F]$ such that $\hat{L}_F(k_F)=k_F\Leftrightarrow L_F(k_F)+k_F=1$ and this $k_F$ is unique.
        \end{enumerate}Observe that if $L_F(p) = p$ (egalitarian
        income distribution), then $k_F = 1/2$. For any other income
        distribution, $1/2 < k_F < 1$. It is interesting to note that
        while the Lorenz curve typically has only two trivial fixed
        points (the two end points), the complementary Lorenz function
        has a unique non-trivial fixed point $k_F$. This fixed point
        $k_F$ lies between 50\% population proportion and the
        population proportion $p^*_F=L_F^{-1}(1/2)$ that we associate
        with 50\% income given the income distribution $F$.




\begin{case}\rm  Let $F$ be the uniform distribution on $[a, b]$ where
  $0 \leq a < b < \infty$.  Then,
  $L_F(p) = p\left[1 - \left\{(1 - p)(b - a)/(b + a)\right\}\right]$
  and
  $$k_F = \frac{-(3a + b) + \sqrt{5a^2 + 6ab + 5b^2}}{2(b - a)}.$$

  It is interesting to note that if $a = 0$, then $L_F(p) = p^2$ and $k_F$ is \emph{the reciprocal of the Golden ratio}, that is, 
  $k_F = (\sqrt{5}-1)/2 = 1/\phi$ where $\phi = (\sqrt{5}+1)/2$ is the
  \emph{Golden ratio}.
 \end{case}

 \begin{case}
 \rm  Let $F$ be the exponential distribution function given by
   $F(x) = 1 - e^{-\lambda x}$ where $x \geq 0$ and $\lambda > 0$.
   Then,
   $L_F(p) = p - (1 - p)\ln (1/(1 - p)), \hat{L}_F(p) = (1 - p)
   \left[1 + \ln\left\{1/(1 - p)\right\}\right]$
   and $k_F \sim 0.6822$.
    \end{case}

\begin{case}
\rm   The Pareto distribution is given by $F(x) = 1 - (x_m/x)^{\alpha}$
   on the support $[x_m,\infty)$ where $\alpha > 1$ and the minimum
   income is $x_m > 0$. Then,
   $L_F(p) = 1 - (1 - p)^{1 - \frac{1}{\alpha}}$ and
   $\hat{L}_F(p) = (1 - p)^{1 - \frac{1}{\alpha}}$.  The $k$-index is
   a solution to $(1 - k_F)^{1 - \frac{1}{\alpha}} = k_F$.
   If $\alpha = \ln 5/\ln 4\sim 1.16$, then $k_F = 0.8$
   and we get what is known as {\it the Pareto principle or the $80/20$ rule}.
 \end{case}

  \subsection{$k$-index as a generalization of the Pareto principle}
  The Pareto principle is based on Pareto's observation (in the year
  1906) that approximately $80$\% of the land in Italy was owned by
  $20$\% of the population.  The evidence, though, suggests that the
  income distribution of many countries fails to satisfy the 80/20
  rule (see \cite{GC2014}).  The $k$-index can be thought of as a
  generalization of the Pareto principle. Note that
  $L_F(k_F) = 1 - k_F$; hence, the top $100(1 - k_F)$\% of the
  population has $100(1 - (1 - k_F)) = 100k_F$\% of the income.
  Hence, the ``Pareto ratio'' for the $k$-index is $k_F/(1 -
  k_F)$. Observe, however, that this ratio is obtained endogenously
  from the income distribution and in general, there is no reason to
  expect that this ratio will coincide with the Pareto principle.\footnote{The fact that the $k$-index generalizes Pareto's 80/20 rule was first pointed out in \cite{GC2014} and later also in \cite{EL2015A}, \cite{EL2016}.}
  
 Given any income distribution $F$, for any $p \in [0, p^*_F]$ with $p^*_F=L_F^{-1}(1/2)$, let $r_F(p) = L^{-1}_F(1 - p)$. Consider the
  interval $C(p) = [\min \{p, r_F(p)\}, \max \{p, r_F(p)\}]$.  To understand
  what $C(p)$ signifies, let $p = 0.2$.  That is, we consider the poorest
  $20\%$ (or $100p\%$) as undeniably poor.  To identify the dividing line between the
  poor and the rich, one strategy is to eliminate those who are
  undeniably rich.  We do this by considering the fraction of the rich
  whose income share is exactly $100(1-p)\%$.  That is, we identify $p'$ such
  that $L_F(p') = 0.8=1-p$. Eliminating the poor and the undeniably rich, we find that the dividing line between poor and rich must lie in the interval $[\min \{p, r_F(p)\}, \max \{p, r_F(p)\}]$. We now ask the question: What proportions are in the set $[\min\{p, r_F(p)\}, \max\{p, r_F(p)\}]$ for all $p\in [0,p^*_F]$?  The answer is that only $k_F$ meets this criterion. Specifically, for any $p\in [0,p^*_F]$, define the potential income disparity division set as $\mathcal{C}(p) = \{t | \min\{p, r_F(p) \} \leq t \leq \max \{p,r_F(p)\}\}$.  We show in the appendix that
  
  \begin{equation}\label{interval}
  k_F = \cap_{p \in [0,p^*_F]}\mathcal{C}(p).
  \end{equation}

        \subsection{Interpreting the $k$-index in terms of rich-poor
          disparity} The \emph{Gini coefficient}, as is well-known,
        measures inequality by the area between the Lorenz curve and
        the $45$-degree line.  For any $p\in [0,1]$, we can decompose
        this coefficient into three parts: two representing the
        \emph{within-group inequality} and one representing the
        \emph{across-group inequality}.  In Figure \ref{fig:disparity}
        below, the unshaded area bounded by the Lorenz curve and the
        line from $(0, 0)$ to $(p, L_F(p))$ is the within-group
        inequality of the poor. It represents the extent to which
        inequality can be reduced by redistributing incomes among the
        poor.  Similarly, the area bounded by the Lorenz curve and the
        line segment from $(p, L_F(p))$ to $(1, 1)$ represents the
        within-group inequality of the rich.  The shaded area
        represents the across-group inequality.

        An easy computation shows that the extent of across-group
        inequality between the bottom $p\times100\%$ and top
        $(1-p)\times 100\%$ is the (across-group) disparity function
        $D_F(p) = (1/2)[p - L_F(p)]$.  One can ask for what value of
        $p$ is the across-group inequality maximized?  The answer is
        that this is maximized at the proportion associated with the
        \emph{Pietra index}.  It is well-known that the Pietra index
        (see \cite{Pi1915}) is given by

$$\mathcal{P}_F: = \max_{p \in [0, 1]}2D_F(p) = \max_{p \in
	[0, 1]}[p - L_F(p)] = F(\mu) - L_F(F(\mu)).$$

\begin{figure}[h]
	\begin{center}
  \begin{tikzpicture}
	\begin{axis}[xlabel={fraction of population}, 
	ylabel={fraction of income},
	xmin=0,
	ymin=0,
	xmax=1,
	ymax=1,
	legend style={
		at={(0.25,0.95)},
		anchor=north,
              }]
	\addplot[color=red, thick, domain=0:1] {x*x};
	\legend{Lorenz curve};
	\addplot[domain=0:1] {x};
	\addplot coordinates
	{(0, 0) (0.5, 0.25) (1, 1)} --cycle [fill=gray];
	\node[right] at (0.51, 0.25){\small $(P, L(P))$};
	\end{axis}
	\end{tikzpicture}
        \caption{The shaded area represents the inter-group
          inequality. \label{fig:disparity}}
      \end{center}
\end{figure}
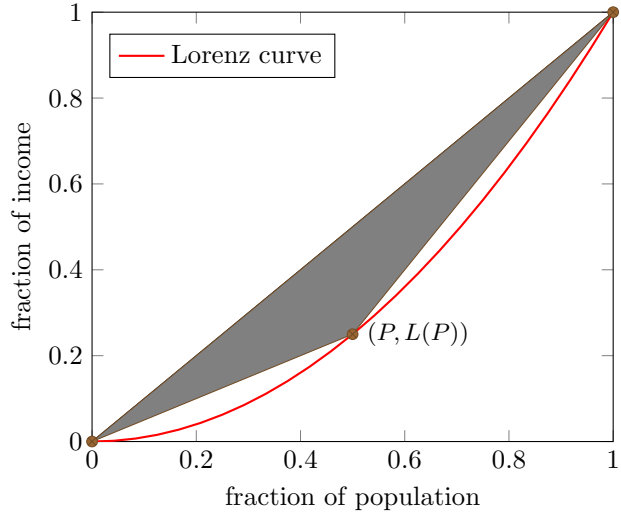

Hence, $F(\mu)$ is the proportion where the disparity is maximized.
Therefore, one way of understanding the Pietra index is that it splits
society into two groups in a way such that inter-group inequality is
maximized.  This provides a different perspective on the Pietra index.

What about the $k$-index?  Let us divide society into two groups, the
``poorest'' who constitute a fraction $p$ of the population and the
``rich'' who constitute a fraction $1 - p$ of the population.  Given
the Lorenz curve $L_F(p)$, we look at the distance of the ``boundary
person'' from the poorest person on the one hand and the distance of
this person from the richest person on the other hand.  These
distances are given by $\sqrt{p^2 + L_F(p)^2}$ and
$\sqrt{(1 - p)^2 + (1 - L_F(p))^2}$ respectively.  Then, the $k$-index
divides society into two groups in a manner such that the Euclidean
distance of the boundary person from the poorest person is equal to
the distance from the richest person.

The value of the disparity function at the $k$-index is given by
$D_F(k_F) = k_F - 1/2$. The interpretation of this is quite
transparent since it measures the gap between the proportion $k_F$ of
the poor from the $50-50$ population split.  As long as we do not have
a completely egalitarian society, $k_F > 1/2$ and hence it is one way
of highlighting the rich-poor disparity with $k_F$ defining the income
proportion of the top $(1-k_F)$ proportion of the rich population.
The other measures do not have as nice an interpretation.  For
instance, the value of the disparity function at the proportion
corresponding to the Pietra index is
$D_F(F(\mu)) = [F(\mu) - L_F(F(\mu))]/2$. This number has no obvious interpretation.

\subsection{The $k$-index as a solution to optimization problems}
The $k$-index is the unique solution to the following surplus
maximization problem:

\begin{equation}\label{one}
k_F = \underset{P\in [0,1]}{\arg\!\max}\int\limits_{0}^{P}(\hat{L}_F(t)-t)dt.
\end{equation}

Therefore, $k_F$ is that fraction of the lower income population for
which the area between the complementary Lorenz function and the
income distribution line associated with the egalitarian distribution
is maximized. Condition (\ref{one}) follows since $\hat{L}_F(p)\geq p$ for all $p\in [0,k_F]$ and $\hat{L}_F(p)<p$ for all $p\in (k_F,1]$. For the same reason the $k$-index is also the unique solution to the
following surplus minimization problem (which is the dual of the
problem in (\ref{one})):

\begin{equation}
k_F = \underset{P\in [0,1]}{\arg\!\min}\int\limits_{P}^{1}\{(1 - t) - L_F(t)\}dt.
\end{equation}

Therefore, $(1-k_F)$ is that fraction of the higher income population
for which the area between the income distribution line associated
with the egalitarian distribution and the Lorenz function is
minimized.
\section{\textbf{Comparing the normalized $k$-index with the Pietra index and the Gini coefficient}} We start our comparison  by specifying a family of Lorenz functions for each of which the Gini coefficient coincides with the Pietra index and yet the normalized $k$-index is different.

\begin{case}
	\rm \label{verynew1}
Consider the $p$-oligarchy Lorenz function discussed in \cite{ES2010} that has the following functional form: For any fraction $a\in (0,1)$, 
\begin{equation}
L_{F}(p)=
\left\{ \begin{array}{ll}
0  & \mbox{if $p\in [0,a]$,} \\
\frac{(p-a)}{(1-a)}  & \mbox{if $p\in (a,1]$.} 
\end{array}
\right.  \label{LF0}
\end{equation}See Figure \ref{GPC} where the Lorenz function given by (\ref{LF0}) is represented by the piecewise linear red line OBA (colour online). 
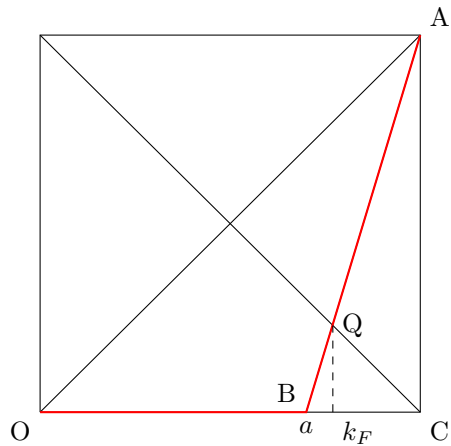
\begin{figure}[h]
	\begin{center}
		\begin{tikzpicture}[scale=5]
                  \draw (0,0)--(0,1)--(1,1)--(1,0)--(0,0); \draw
                  [white](0,0)--(0.7,0); \draw (0,0)--(1,1) node
                  [pos=0,below left] {O} node [pos=1,above right] {A};
                  \draw (0,1)--(1,0) node [pos=1,below right] {C};
                  \draw (0.7,0)--(1,1) node[pos=0,below, black] {$a$}
                  node[pos=0,above left,thick,black]{B};
                  \draw [thick,red](0,0)--(0.7,0)--(1,1); \draw
                  [dashed](0.77,0)--(0.77,0.23) node[pos=0,below
                  right]{$k_F$} node[pos=1,above,right]{Q};
		\end{tikzpicture}
	\end{center}
	\caption{$G_F=\mathcal{P}_F=a>\mathcal{K}_F$. \label{GPC}}
	\end{figure}It is easy to verify that the proportion associated with the  Pietra index is $F(\mu)=a$ and
        $\mathcal{P}_F=F(\mu)-L_F(F(\mu))=a-0=a$. One can also verify
        that the Gini coefficient coincides with the Pietra index,
        that is, $G_{F}=\mathcal{P}_F=a$. However, the $k$-index
        fraction $k_F$ is a solution to the equation
        $(k_F-a)/(1-a)+k_F=1$ and it gives $k_F=1/(2-a)$. Moreover, the
        normalized $k$-index yields
        $\mathcal{K}_F=2k_F-1=a/(2-a)$. Therefore, for any given
        $a\in (0,1)$ and any associated Lorenz function given by
        (\ref{LF0}), we have
\begin{equation}
G_F=\mathcal{P}_F=a>\frac{a}{2-a}=\mathcal{K}_F. 
\end{equation}
\end{case} Therefore, Case \ref{verynew1} suggests that the $k$-index in itself has properties that are different from the other two measures and hence deserves a special theoretical analysis.
\subsection{Coincidence of the $k$-index and the Pietra index}
The Lorenz function $L_F(p)$ is \emph{symmetric} if for all $p\in [0,1]$,
\begin{equation}
\label{kpietra}
L_F(\hat{L}_F(p))=1-p \ {\rm or} \ {\rm equivalently} \ L_F(p)+r_F(p)=1,
\end{equation}where $r_F(p)=L^{-1}_F(1-p)$. The idea of symmetry is explained in Figure \ref{symmetryLorenzF}.

\begin{figure}[h]
	\begin{center}
		\begin{tikzpicture}[scale=5]
		\draw (0,0)--(0,1)--(1,1)--(1,0)--(0,0);
		\draw (0,0)--(1,1) node [pos=0,below left] {O}  ;
		\draw (0,1)--(1,0) ;
		\draw [red,thick](0,0)..controls(0.75,0.25)..(1,1);
		\draw [blue,thick](0.22,0)--(0.22,0.075)node[pos=0,below,black]{B} node[pos=1,above right,black]{A};
		\draw [dashed](0.22,0.075)--(0.22,0.781);
		\draw [dashed](0,0.781)--(0.925,0.781);
		\draw [blue,thick](0.925,0.781)--(1,0.781)node[pos=0,above left,black]{C} node[pos=1, right,black]{D};
		\end{tikzpicture}
		\caption{Symmetry condition (\ref{kpietra}) requires that for any proportion $p$, that the distance $L_F(p)$ between the points $A=(p,L_F(p))$ and $B=(p,0)$ must be the same as the distance $1-L^{-1}_F(1-p)$ between the points $C=(L^{-1}_F(1-p), (1-p))$ and $D(1,1-p)$.  \label{symmetryLorenzF}}
	\end{center}
\end{figure}
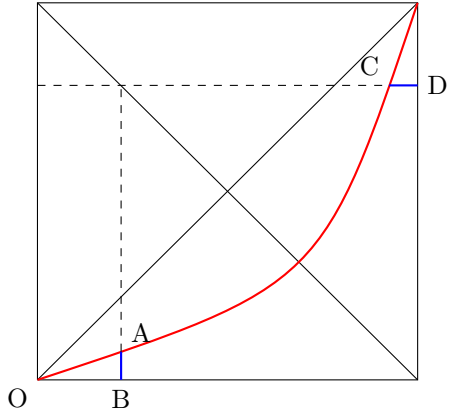
\begin{case}\rm \label{CKPNG} Suppose that the Lorenz function is given by $L_F(p)=1-\sqrt{1-p^2}$ (see Figure \ref{PKC}). Observe that $L_F(\hat{L}_F(p))=L_F(\sqrt{1-p^2})=1-p$ and hence the Lorenz function $L_F(p)=1-\sqrt{1-p^2}$ is symmetric.
	
	 \begin{figure}[h]
	 	\begin{center}
		\begin{tikzpicture}[scale=3]
		
		\draw [red](1,1)circle(1cm);
		\draw (1,1)--(1,0) node [pos=0,above ] {C$=(0,1)$}  node [pos=1,below ] {O};
		\draw (1,1)--(2,1)   node [pos=1, right] {B};
		\draw (2,0)--(2,1) node [pos=0,below ] {A} ;
		\draw (1,0)--(2,0);
		\draw (1,0)--(2,1);
		\draw [dashed](1,1)--(2,0);
		\draw [dashed](1.71,0)--(1.71,0.29) node [pos=0,below] {$k_F$}  node [pos=1,above] {D};
		\end{tikzpicture}
		\caption{$\mathcal{K}_F=\mathcal{P}_F=\sqrt{2}-1<G_F=\pi/2-1$.  \label{PKC}}
		\end{center}
	\end{figure}
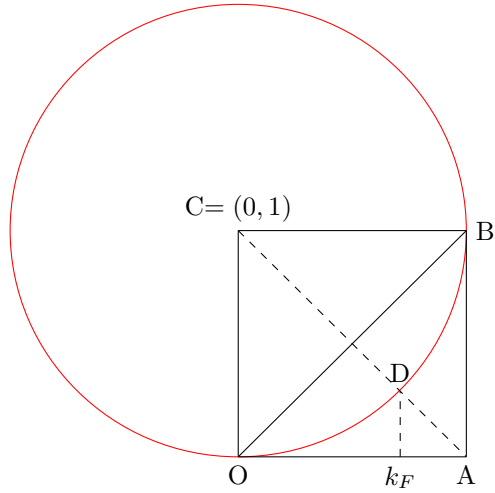
	
The $k$-index associated with the Lorenz function $L_F(p)=1-\sqrt{1-p^2}$ is $k_F=1/\sqrt{2}$. Moreover, since $L'_F(p)=p/\sqrt{1-p^2}$, at the proportion $F(\mu)$ associated with the Pietra index $\mathcal{P}_F$, we have $L'_F(F(\mu))=F(\mu)/\sqrt{1-\{F(\mu)\}^2}=1$ implying $F(\mu)=k_F$. Therefore, for the Lorenz function given by $L_F(p)=1-\sqrt{1-p^2}$, the normalized $k$-index $\mathcal{K}_F=2k_F-1=\sqrt{2}-1$ coincides with the Pietra index $\mathcal{P}_F=F(\mu)-L_F(F(\mu))=\sqrt{2}-1$.  Moreover, one can verify that the Gini coefficient is different and is given by $G_F=\pi/2-1>\mathcal{P}_F=\mathcal{K}_F$. 
\end{case} Case \ref{CKPNG} provides an example of a symmetric and differentiable Lorenz function for which $k_F=F(\mu)$ and hence $\mathcal{K}_F=\mathcal{P}_F$. This result is true in general and in the appendix we prove the following general result.  
\begin{enumerate}
\item[{\bf (KP)}]  If the Lorenz function is symmetric and differentiable, then the proportion $F(\mu)$ associated with the Pietra index coincides with the proportion $k_F$ of the $k$-index. Hence, we also have $\mathcal{K}_F=\mathcal{P}_F$
\end{enumerate}Observe that {\bf (KP)} provides a sufficient condition for the coincidence. It is not necessary as the following example shows that we can find a Lorenz function which is not symmetric and yet we have the coincidence of the normalized $k$ and the Pietra indices. 

\begin{equation}
L_{F}(p)=
\left\{ \begin{array}{ll}
1-\sqrt{1-p^2}  & \mbox{if $p\in [0,1/\sqrt{2}]$,} \\
1-\frac{\sqrt{3}}{2-\sqrt{2}}(1-p)  & \mbox{otherwise.} 
\end{array}
\right.  \label{LF10}
\end{equation}Note that in (\ref{LF10}) we have simply replaced the curve DB in Figure \ref{PKC} by a straight line between the two points. Even though this Lorenz curve is not symmetric, we can ``convert'' it into a symmetric one by replacing the segment OD by a corresponding straight line.  This change leaves $\mathcal{K}_F$ and $\mathcal{P}_F$ unchanged.  It is clear  that this can be done in general: given any non-symmetric Lorenz curve where $\mathcal{K}_F$ and $\mathcal{P}_F$ coincide, we can derive a symmetric Lorenz curve such that the two indices coincide by replacing the Lorenz curves for the poor and the rich by straight lines.  This suggests that the symmetry condition is almost necessary.

\subsection{Coincidence of the normalized $k$-index and the Gini coefficient}  As an instance of coincidence between $G_F$ and $\mathcal{K}_F$ we consider the following family of Lorenz functions. 

\begin{case} \rm \label{allcoincidence} For any fraction $K\in [1/2,1)$, consider the associated Lorenz function $L_F(p)$ given by
\begin{equation}\label{LFGKP}
L_{F}(p)=
\left\{ \begin{array}{ll}
\left(\frac{1-K}{K}\right)p  & \mbox{if $p\in [0,K]$,} \\
(1-K)+\frac{K}{(1-K)}(p-K)  & \mbox{if $p\in(K,1]$.} 
\end{array}
\right.  
\end{equation}In Figure \ref{LFCKGP}, the Lorenz function given by (\ref{LFGKP}) is depicted by the piecewise linear red lines OQ and QB (colour online).  
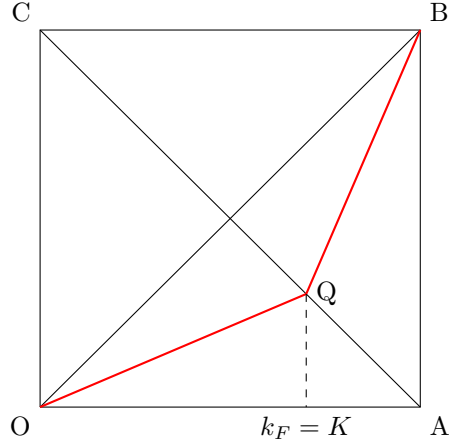
\begin{figure}[h]
	\begin{center}
		\begin{tikzpicture}[scale=5]
                  \draw (0,0)--(0,1)--(1,1)--(1,0)--(0,0); \draw
                  (0,0)--(1,1) node [pos=0,below left] {O} node
                  [pos=1,above right] {B}; \draw (0,1)--(1,0)node
                  [pos=0,above left] {C} node [pos=1,below right] {A};
                  \draw [red,thick](0,0)--(0.7,0.3) node [pos=1,right,
                  black] {Q}; \draw [red,thick](0.7,0.3)--(1,1); \draw
                  [dashed](0.7,0.3)--(0.7,0) node
                  [pos=1,below]{$k_F=K$};
		\end{tikzpicture}
	\end{center}
	\caption{$\mathcal{K}_F=G_F=2K-1=\mathcal{P}_F$. \label{LFCKGP}}
\end{figure}

It is immediate that $L_F(K)+K=(1-K)+K=1$ implying that $k_F=K$. Moreover, $\int_{t=0}^{t=1}L_F(t)dt=1-K$ and hence $G_F=1-2(1-K)=2K-1=2k_F-1=\mathcal{K}_F$. Also note that the difference $p-L_F(p)$ is maximized at $p=K$ and hence $F(\mu)=k_F=K$. Therefore, we have  
\begin{equation}
G_F=\mathcal{K}_F=\mathcal{P}_F=2K-1.\footnote{Observe that if $K=1/2$, then from (\ref{LFGKP}) we have the Lorenz function for egalitarian income, that is, $L_F(p)=p$ for all $p\in [0,1]$ and in that case $G_F=\mathcal{K}_F=\mathcal{P}_F=0$.}
\end{equation}
\end{case} Therefore, Case \ref{allcoincidence} shows that for the family of Lorenz functions given by (\ref{LFGKP}), $G_F$ coincidences with $\mathcal{K}_F$ and as a result  $\mathcal{P}_F$ also coincides. We claim that this is no exception. More generally, in the appendix we show the following:

\begin{enumerate}
	\item[{\bf (GK-P)}] For any income distribution $F$, $G_F\geq \mathcal{P}_F\geq \mathcal{K}_F$. Moreover, if $G_F=\mathcal{P}_F=\mathcal{K}_F$, then the Lorenz function is given by (\ref{LFGKP}).
\end{enumerate} Eliazar \cite{EL2015A} obtains the same order across the three indices using appropriate maximization exercises on the Lorenz set which is obtained by taking the area between two types of Lorenz curves on the unit square. The first type is the standard Lorenz curve $L_F(p)$ and the second type of Lorenz curve is  $\bar{L}_F(p)=(1/\mu)\int_{t=1-p}^{t=1}F^{-1}(t)dt$. However, the technique we apply to obtain the order across the three indices is mainly based on convexity of the Lorenz function (see Appendix). Moreover, we not only provide the order across the three indices, we also identify the complete family of Lorenz functions for which the three indices coincide.

\section{\textbf{Ranking the Lorenz functions using the normalized $k$-index, the Pietra index and the Gini coefficient}} One important aspect of summary statistics is to rank different Lorenz curves. Here we demonstrate that the three indices can provide very different rankings.  
\begin{case}
\rm 	Consider the following Lorenz functions: 
\begin{equation}
L_{F_1}(p)=
\left\{ \begin{array}{ll}
\frac{3p}{4}  & \mbox{if $p\in [0,1/3]$,} \\
\frac{9p-1}{8}   & \mbox{if $p\in (1/3,1]$.} 
\end{array}
\right.  \label{LF1}
\end{equation} 

\begin{equation}
L_{F_2}(p)=
\left\{ \begin{array}{ll}
\frac{8p}{9}  & \mbox{if $p\in [0,7/8]$,} \\
\frac{16p-7}{9}  & \mbox{if $p\in (7/8,1]$.} 
\end{array}
\right.  \label{LF2}
\end{equation}Observe that the population fraction associated with the Pietra index is $F_1(\mu_1)=1/3$ for the Lorenz function $L_{F_1}(p)$ and is $F_2(\mu_2)=7/8$ for the Lorenz function $L_{F_2}(p)$. Since $\mathcal{P}_{F_i}=F_i(\mu_i)-L_{F_i}(F_i(\mu_i))$ for $i=1,2$, we get   
\begin{equation}
\mathcal{P}_{F_1}=\frac{1}{12}<\mathcal{P}_{F_2}=\frac{7}{72}.
\end{equation}The $k$-index fraction $k_{F_1}$ associated with the Lorenz function $L_{F_1}(p)$ is a solution to the equation $(9k_F-1)/8+k_{F_1}=1$ and it gives $k_{F_1}=9/17$. The $k$-index fraction $k_{F_2}$ associated with the Lorenz function $L_{F_2}(p)$ is a solution to the equation $8k_{F_2}/9+k_{F_2}=1$ and it also gives $k_{F_2}=9/17$. Therefore, $k_{F_1}=k_{F_2}=9/17$ and hence the normalized $k$-indices are also identical, and, in particular, we have 
\begin{equation}\label{rankkp}
\mathcal{K}_{F_1}=\mathcal{K}_{F_2}=\frac{1}{17}<\mathcal{P}_{F_1}=\frac{1}{12}<\mathcal{P}_{F_2}=\frac{7}{72}.
\end{equation}
\end{case}

\begin{case}\label{policy} \rm Now consider the following two Lorenz functions: 
\begin{equation}
L_{F_3}(p)=p^2, \ \ \forall \ \ p\in [0,1].  \label{LFkg1}
\end{equation} 

\begin{equation}
L_{F_4}(p)=
\left\{ \begin{array}{ll}
p^2  & \mbox{if $p\in [0,3/4]$,} \\
1-\frac{7(1-p)}{4}  & \mbox{if $p\in (3/4,1]$.} 
\end{array}
\right.  \label{LFkg2}
\end{equation}The $k$-index associated with both Lorenz functions $L_{F_3}(p)$ and $L_{F_4}(p)$ is a solution to the equation $K^2+K=1$ and it gives $k_{F_3}=k_{F_4}=K=1/\phi$ where $\phi=(\sqrt{5}+1)/2$ is the Golden ratio. Therefore, $\mathcal{K}_{F_3}=\mathcal{K}_{F_4}=2/\phi-1\simeq 0.23607$. However, Gini coefficient associated with the two Lorenz functions $L_{F_3}(p)$ and $L_{F_4}(p)$ are different. In particular, one can show that $G_{F_3}=2\int_{0}^1[t-t^2]dt=1/3$ and $G_{F_4}=2\int_{0}^{3/4}[t-t^2]dt+\int_{3/4}^1[(3/4)(1-t)]dt=21/64$.
\begin{equation}\label{rankkp}
\mathcal{K}_{F_3}=\mathcal{K}_{F_4}=2/\phi-1<G_{F_4}=21/64<G_{F_3}=1/3.
\end{equation}
\end{case}Case \ref{policy} demonstrates an important difference between $\mathcal{K}_F$ and $G_F$. The Gini is affected by transfers within a group. In particular, the poor are unaffected but the rich have become more egalitarian while moving from $L_{F_3}$  to $L_{F_4}$. The normalized $k$-index on the other hand is unaffected with such intra-group transfers. This suggests that if we are interested in reducing inequality between groups, then the normalized $k$-index is a better indicator.   

\section{Summary} We summarize the main results of this paper: 
\begin{enumerate}
	\item The $k$-index always exists and is a unique fixed point of the complementary Lorenz function. While the Lorenz function has two trivial fixed points, the complementary Lorenz function has one non-trivial fixed point $k_F$ and it gives the value of the Kolkata index or the $k$-index (see Section 3.1).
	\item The $k$-index generalizes Pareto's $80/20$ rule. The $k$-index has the property that $[100(1 - k_F)]$\% of the people own $[100k_F]$\% of the income.  We also provide an argument as to why $k_F$ is a correct and endogenously obtained dividing population proportion between the rich and the poor in a society with income distribution $F$ (see condition (\ref{interval}) on Section 3.2).   
\item Although the $k$ and Pietra indices both split the society into two groups,  the $k$-index is more transparent measure for the poor-rich split. 
\item The $k$-index also has interpretations as a solution to optimization problems. The $k$-index maximizes the area between the complementary Lorenz function and the income distribution line associated with the egalitarian distribution.  Hence, $(1-k_F)$ minimizes the area between the income distribution line associated with the egalitarian distribution and the Lorenz function. 
\item We compare the normalized $k$-index ($\mathcal{K}_F:=2k_F-1$) with the Gini coefficient $G_F$ and the Pietra index $\mathcal{P}_F$. If the Lorenz function is symmetric, then the normalized $k$-index coincides with the Pietra index (see Section 4.1).  We show for any given income distribution, $G_F\geq \mathcal{P}_F\geq \mathcal{K}_F$. We have also identified the complete set of Lorenz functions for which the coincidence between the normalized $k$-index with the Gini coefficient and the Pietra index takes place (see Section 4.2).   

\item Finally, we show that the ranking of Lorenz functions from the $k$-index is different from that of the Pietra index as well as from the Gini coefficient. The Gini coefficient and the Pietra index are affected by transfers exclusively among the rich or among the poor, the $k$-index ranks is only affected by transfers across the two groups (see Section 5). 
\end{enumerate}We conclude by noting that throughout our paper we have specified the $k$-index, the Gini coefficient and the Pietra index as measures of income inequality. Clearly, income inequality is just one application of the ideas in this paper. We can use these indices for measuring wealth or other social inequalities. In general, the $k$-index can be useful for quantifying heterogeneity in any social system.  

\section{Appendix}

\vspace{.1in}

\noindent
{\bf Proof of (\ref{interval}):}	If $p= k_F$, then $\min\{k_F,r_F(k_F)\} = \max\{k_F,r_F(k_F)\}=k_F$ implying  $\mathcal{C}(k_F)=\{k_F\}$. If $p\in [0, k_F)$, then $L_F(k_F) = 1 - k_F < 1 - p$ and from non-decreasingness of $L_F(.)$ we get $k_F\leq r(p)$.	Therefore, if $p\in [0, k_F)$, then $p < k_F \leq r_F(p)$ and $k_F\in \mathcal{C}(p)$. Similarly, if $p\in (k_F,p^*_F]$, then $L_F(k_F) = 1 - k_F > 1 - p\Rightarrow k_F \geq r_F(p)$. Therefore, if $p\in (k_F,p^*_F]$, then $p> k_F \geq r_F(p)$ and $k_F\in \mathcal{C}(p)$.  \qed

\vspace{.1in}

\noindent
{\bf Proof of (KP):} Specifically, using the symmetry and differentiability of the Lorenz function it follows that $-(1/L'_F(r_F(p)))+L'_F(p)=0$ and, given $L'_F(F(\mu))=1$ at the population fraction $F(\mu)$ associated with Pietra index, it follows that $L'_F(r_F(F(\mu)))=1\Rightarrow F^{-1}(r_F(F(\mu)))=\mu\Rightarrow L^{-1}_F(1-F(\mu)))=F(\mu)\Rightarrow F(\mu)+L_F(F(\mu))=1$ implying $F(\mu)=k_F$. \qed

\vspace{.1in}

\begin{figure}[h]
	\begin{center}
		\begin{tikzpicture}[scale=5]
		\draw (0,0)--(0,1)--(1,1)--(1,0)--(0,0);
		\draw (0,0)--(1,1) node [pos=0,below left] {O}  node [pos=1,above right] {B};
		\draw (0,1)--(1,0) ;
		\draw (0.688,0)--(0.688,0.313);
		\draw [red,thick](0,0)..controls(0.75,0.25)..(1,1);
		\draw [thick](0,0)--(0,0.313);
		\draw [blue,dashed,thick](0,0)--(0.688,0.313) node[pos=1,right,black]{A};
		\draw [blue,dashed,thick](0.688,0.313)--(1,1);
		\draw [decorate,decoration={brace,amplitude=10pt},xshift=0pt,yshift=0pt]
		(0,0) -- (0.688,0)node [pos=0.5,above right,xshift=0pt,yshift=6pt] {$q$};
		\draw [decorate,decoration={brace,amplitude=10pt},xshift=0pt,yshift=0pt]
		(0.688,0)--(0.688,0.313)node[pos=0.5,below right,xshift=-4pt,yshift=6pt]{$L_F(q)$};
		\end{tikzpicture}
		\caption{The red curve OAB depicts any Lorenz function $L_F(p)$ and, for any $q\in (0,1)$, the dotted piecewise linear blue line OAB is the induced Lorenz function $\bar{L}_F(p)$ (colour online). \label{LFKG-general}}
	\end{center}
\end{figure}
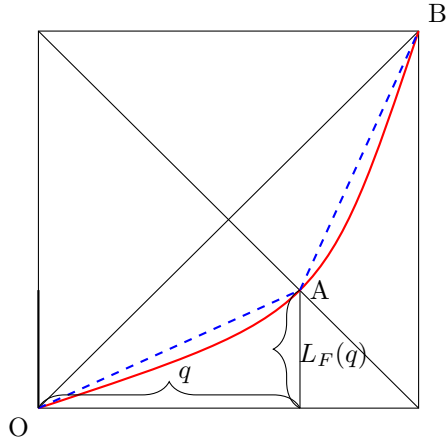

\noindent
\noindent
{\bf Proof of {\bf (GK-P)}:} Consider any Lorenz function $L_F(p)$ and for any $q\in (0,1)$ define the induced Lorenz function 
\begin{equation}
\bar{L}_{F}(p)=
\left\{ \begin{array}{ll}
\frac{L_F(q)}{q}p  & \mbox{if $p\in [0,q]$,} \\
\frac{(1-L_F(q))}{(1-q)}p-\frac{(q-L_F(q))}{(1-q)}  & \mbox{if $p\in (q,1]$.} 
\end{array}
\right.  \label{LPKCOrdefr}
\end{equation}
In Figure \ref{LFKG-general}, we depict how given any $q\in (0,1)$ we get the induced Lorenz function $\bar{L}_F(p)$ from any given Lorenz function $L_F(p)$.  

From Figure \ref{LFKG-general}, it is clear that $\bar{L}_F(p)\geq L_F(p)$ for all $p\in [0,1]$. Hence, \[ \begin{split}\int_{0}^{1}L_F(t)dt & \leq \int_{0}^{q}\bar{L}_F(t)dt+\int_{q}^{1}\bar{L}_F(t)dt \\ & =\int_{0}^{q}\frac{L_F(q)}{q}tdt+\int_{q}^{1}\left\{\frac{(1-L_F(q))}{(1-q)}t-\frac{(q-L_F(q))}{(1-q)}\right\}dt \\ & =\frac{qL_F(q)}{2}+L_F(q)-q+\frac{(1-L_F(q))(1+q)}{2} \\ & =\frac{1+L_F(q)-q}{2}.\end{split} \]Since $\int_{t=0}^{t=1}L_F(t)dt=(1-G_F)/2$, it follows that $G_F\geq q-L_F(q)$ for any $q\in [0,1]$. Since Pietra index maximizes the function $q-L_F(q)$ over all $q\in [0,1]$, it follows that $G_F\geq F(\mu)-L_F(F(\mu))\geq k_F-L_F(k_F)$ implying $G_F\geq \mathcal{P}_F\geq \mathcal{K}_F$.

We now show that if the Gini coefficient coincides with the normalized $k$ index, then the Lorenz function must be given by (\ref{LFGKP}). Consider any population proportion $K\in [1/2,1)$ and given such a $K$ consider any income distribution $F$ such that $k_F=K$.\footnote{Possibility of such a selection is guaranteed by the family of Lorenz functions defined by (\ref{LFGKP}).}  Then, the Gini coefficient $G_F=1-2\int_{t=0}^{t=1}L_F(t)dt$ coincides with the normalized $k$-index $\mathcal{K}_F=K-L_F(K)$ if and only if 
\begin{equation}\label{cond0}
\int\limits_{t=0}^{t=1}L_F(t)dt=L_F(K)\Leftrightarrow \int\limits_{t=0}^{t=K}\{L_F(K)-L_F(t)\}dt=\int\limits_{t=K}^{t=1}\{L_F(t)-L_F(K)\}dt.
\end{equation}

\begin{figure}[h]
	\begin{center}
		\begin{tikzpicture}[scale=5]
		\draw (0,0)--(0,1)--(1,1)--(1,0)--(0,0);
		\draw (0,0)--(1,1) node [pos=0,below left] {O}  node [pos=1,above right] {D};
		\draw (0,1)--(1,0) ;
		\draw [red,thick](0,0)..controls(0.75,0.25)..(1,1);
		\draw [red,thick](0,0.313)--(0.688,0.313)  node [pos=0,left,black]{B} node [pos=1,below,black]{A};
		\draw [red,thick](0.688,0.313)--(1,0.313) node [pos=1,right,black]{C};
		\draw [red,thick](0,0)--(0,0.313);
		\draw [red,thick](1,1)--(1,0.313);
		\draw [dashed,thick](0,0)--(0.688,0.313);
		\draw [dashed,thick](0.688,0.313)--(1,1);
		\draw [decorate,decoration={brace,amplitude=10pt},xshift=0pt,yshift=0pt]
		(0,0.313) -- (0.688,0.313)node [pos=0.5,above,xshift=0pt,yshift=6pt] {k};
		\draw [decorate,decoration={brace,amplitude=10pt},xshift=0pt,yshift=0pt]
		(0.688,0.313)--(1,0.313) node [pos=0.5,below,xshift=0pt,yshift=2pt] {(1-k)};
		\draw [decorate,decoration={brace,amplitude=10pt},xshift=0pt,yshift=0pt]
		(1,0.313)--(1,1)node [pos=0.5,below right,xshift=0pt,yshift=6pt] {k};
		\draw [decorate,decoration={brace,amplitude=10pt},xshift=0pt,yshift=0pt]
		(0,0)--(0,0.313)node[pos=0.5,left,xshift=-4pt,yshift=6pt]{(1-k)};
		\end{tikzpicture}
			\caption{$\mathcal{K}_F=G_F=\mathcal{P}_F=2K-1$. \label{LFKG-p}}
	\end{center}
\end{figure}

Consider Figure \ref{LFKG-p} where the area of integral $\int_{t=0}^{t=K}\{L_F(K)-L_F(t)\}dt$ is depicted by the region OAB. Given convexity of the Lorenz function, the area OAB is minimized if OAB represents the area of a triangle with base length $K$ and altitude length $(1-K)$. Therefore, we have 
\begin{equation}\label{cond1}
\int\limits_{t=0}^{t=K}\{L_F(K)-L_F(t)\}dt\geq \frac{K(1-K)}{2}.
\end{equation} 

Similarly, in Figure \ref{LFKG-p}, the area of integral $\int_{t=K}^{t=1}\{L_F(t)-L_F(K)\}dt$ is depicted by the region ACD. Given convexity of the Lorenz function, the area ACD is maximized if ACD  represents the area of a triangle with base length $(1-K)$ and altitude length $K$. Therefore, we also have 
\begin{equation}\label{cond2}
\int\limits_{t=K}^{t=1}\{L_F(t)-L_F(K)\}dt\leq \frac{K(1-K)}{2}.
\end{equation} From (\ref{cond1}) and (\ref{cond2}) it follows that 
\begin{equation}\label{cond3}
\int\limits_{t=0}^{t=K}\{L_F(K)-L_F(t)\}dt\geq \frac{K(1-K)}{2}\geq \int\limits_{t=K}^{t=1}\{L_F(t)-L_F(K)\}dt.
\end{equation} Applying (\ref{cond3}) in (\ref{cond0}) we get 

\begin{equation}\label{cond4}
\int\limits_{t=0}^{t=K}\{L_F(K)-L_F(t)\}dt=\frac{K(1-K)}{2}=\int\limits_{t=K}^{t=1}\{L_F(t)-L_F(K)\}dt.
\end{equation}Simplification of the first equality in (\ref{cond4}) gives
\begin{equation}\label{cond5}
\int\limits_{t=0}^{t=K}L_F(t)dt=\frac{K(1-K)}{2}=\int\limits_{t=0}^{t=K}H(t)dt, 
\end{equation} where $H(t):=\frac{(1-K)}{K}t$ for all $t\in [0,K]$.\footnote{Note that $\int_{t=0}^{t=K}H(t)dt=\int_{t=0}^{t=K}\frac{(1-K)}{K}tdt=\frac{(1-K)}{2K}\left\{t^2\right\}_{t=0}^{t=K}=\frac{K(1-K)}{2}$.} Observe that $L_F(0)=H(0)=0$ and $L_F(K)=H(K)=1-K$. For any $t\in [0,K]$, $H(t)$ is increasing and linear in $t$ and $L_F(t)$ is non-decreasing and convex in $t$ and hence $L_F(t)\leq H(t)$ for all $t\in [0,K]$. Therefore, given $\int_{t=0}^{t=K}L_F(t)dt=\int_{t=0}^{t=K}H(t)dt$ (condition (\ref{cond5})), we have $L_F(t)=H(t)$ for all $t\in [0,K]$, that is,
\begin{equation} 
L_F(t)=\frac{(1-K)}{K}t, \ \ \ \forall \  t\in [0,K]. \label{cond6}
\end{equation}

\noindent
Similarly, simplification of the second equality in (\ref{cond4}) gives
\begin{equation}\label{cond7}
\int\limits_{t=K}^{t=1}L_F(t)dt=\frac{K(1-K)}{2}+(1-K)^2=\int\limits_{t=0}^{t=K}I(t)dt, 
\end{equation}where $I(t):=(1-K)+\frac{K}{(1-K)}(t-K)$ for all
$t\in [K,1]$.\footnote{Note that
  $\int_{t=K}^{t=1}I(t)dt=\int_{t=K}^{t=1}\left\{(1-K)+\frac{K}{(1-K)}(t-K)\right\}dt=(1-K)^2+\frac{K}{2(1-K)}\left\{(t-K)^2\right\}_{t=K}^{t=1}=(1-K)^2+\frac{K(1-K)}{2}$.}
Observe that $L_F(K)=I(K)=1-K$ and $L_F(1)=I(1)=1$. For any
$t\in [K,1]$, $I(t)$ is increasing and linear in $t$ and $L_F(t)$ is
non-decreasing and convex in $t$ and hence $L_F(t)\leq I(t)$ for all
$t\in [K,1]$. Therefore, given
$\int_{t=K}^{t=1}L_F(t)dt=\int_{t=K}^{t=1}I(t)dt$ (condition
(\ref{cond7})) we get $L_F(t)=I(t)$ for all $t\in [K,1]$, that is,
\begin{equation} 
L_F(t)=(1-K)+\frac{K}{(1-K)}(t-K), \ \ \ \forall \  t\in [K,1]. \label{cond8}
\end{equation}

Therefore, if for any income distribution $F$, the Gini coefficient
$G_F$ coincides with the normalized $k$-index $\mathcal{K}_F$, then
from (\ref{cond6}) and (\ref{cond8}) (and due to the fact that while
selecting any income distribution $F$ such that $k_F=K$, the selection
of $K\in [1/2,1)$ was arbitrary) it follows that the Lorenz function
must be of the form given by (\ref{LFGKP}).   \qed

\vspace{.2in}

\noindent
{\bf Acknowledgments:} The authors are grateful to Satya R. Chakravarty and Subramanian Sreenivasan for helpful comments and suggestions. They are also thankful to Arindam Paul for helping with some figures.


\end{document}